\documentclass[darftclsnofoot, onecolumn]{IEEEtran}

\usepackage[utf8]{inputenc}
\usepackage{graphicx}
\usepackage{mathtools}
\usepackage{cleveref}   
\usepackage{glossaries}
\usepackage{amsfonts}    
\usepackage{caption}
\usepackage{subcaption}
\usepackage{algorithm}    
\usepackage{algorithmic}
\usepackage{fixltx2e}       
\usepackage{cite}           
\usepackage{url}	    

\newcommand{\vect}[1]{\boldsymbol{#1}}      

\newcommand{\sgn}{\operatorname{sgn}}
\newcommand{\nint}{\operatorname{nint}}

\robustify{\gls}	
\robustify{\Gls}	

\makeglossaries
\newacronym{a-bft}{A-BFT}{association beamforming training}
\newacronym{a-msdu}{A-MSDU}{aggregated MSDU}
\newacronym{ack}{ACK}{acknowledgement}
\newacronym{adc}{ADC}{analog-to-digital converter}
\newacronym{agc}{AGC}{automatic gain control}
\newacronym{aoa}{AoA}{angle of arrival}
\newacronym{aod}{AoA}{angle of departure}
\newacronym{api}{API}{application programming interface}
\newacronym{ap}{AP}{access point}
\newacronym{at}{AT}{announcement time}
\newacronym{awv}{AWV}{antenna weight vector}
\newacronym{bc}{BC}{beam combining}
\newacronym{beb}{BEB}{binary exponential backoff}
\newacronym{ber}{BER}{bit error rate}
\newacronym{bf}{BF}{beamforming}
\newacronym{bi}{BI}{beacon interval}
\newacronym{brp}{BRP}{beam refinement protocol}
\newacronym{bss}{BSS}{basic service set}
\newacronym{bst}{BST}{beam switching/steering and tracking}
\newacronym{bti}{BTI}{beacon time interval}
\newacronym{bt}{BT}{beam tracking}
\newacronym{cap}{CAP}{contention access period}
\newacronym{cbap}{CBAP}{contention based access period}
\newacronym{cci}{CCI}{co-channel interference}
\newacronym{cdma}{CDMA}{Code-Division Multiple Access}
\newacronym{ce}{CE}{channel estimation}
\newacronym{csi}{CSI}{channel state information}
\newacronym{csma/ca}{CSMA/CA}{carrier sense multiple access with collision avoidance}
\newacronym{ctap}{CTAP}{channel time allocation period}
\newacronym{cta}{CTA}{channel time allocation}
\newacronym{cts}{CTS}{clear to send}
\newacronym{dac}{DAC}{digital-to-analog converter}
\newacronym{dev}{DEV}{device}
\newacronym{dmg}{DMG}{directional multi-gigabit}
\newacronym{dnrf}{DNRF}{drift and noise removal filter}
\newacronym{dps}{dps}{degrees per second}
\newacronym{dti}{DTI}{data transmission interval}
\newacronym{fcc}{FCC}{federal communications commission}
\newacronym{fst}{FST}{fast session transfer}
\newacronym{gps}{GPS}{global positioning system}
\newacronym{hmm}{HMM}{hidden markov model}
\newacronym{hpbw}{HPBW}{half-power beamwidth}
\newacronym{iss}{ISS}{initiator sector sweep}
\newacronym{knn}{k-NN}{k-nearest neighbors}
\newacronym{los}{LOS}{line of sight}
\newacronym{lpf}{LPF}{low-pass filter}
\newacronym{mac}{MAC}{medium access control}
\newacronym{ma}{MA}{moving average}
\newacronym{mcs}{MCS}{modulation and coding scheme}
\newacronym{mcta}{MCTA}{management channel time allocation}
\newacronym{mems}{MEMS}{microelectromechanical  system}
\newacronym{midc}{MIDC}{multiple sector ID detection capture}
\newacronym{mid}{MID}{multiple sector ID detection}
\newacronym{mimo}{MIMO}{multiple-input and multiple-output}
\newacronym{miso}{MISO}{multiple-input and single-output}
\newacronym{mmse}{MMSE}{minimum mean square error}
\newacronym{mm}{MM}{markov model}
\newacronym{mp}{MP}{matching pursuit}
\newacronym{msdu}{MSDU}{MAC service data unit}
\newacronym{nlos}{NLOS}{non line of sight}
\newacronym{odr}{ODR}{output data rate}
\newacronym{ofdm}{OFDM}{orthogonal frequency division multiplexing}
\newacronym{par}{PAR}{Project Authorization Request}
\newacronym{pbss}{PBSS}{personal basic service set}
\newacronym{pcp}{PCP}{PBSS central point}
\newacronym{per}{PER}{packet error rate}
\newacronym{pet}{PET}{pattern estimation and tracking}
\newacronym{phy}{PHY}{physical}
\newacronym{plcp}{PLCP}{physical layer convergence protocol}
\newacronym{pnc}{PNC}{piconet controller/coordinator}
\newacronym{rf}{RF}{radio frequency}
\newacronym{rpm}{RPM}{rotations per minute}
\newacronym{rps}{RPS}{Radiowave Propagation Simulator}
\newacronym{rssi}{RSSI}{received signal strength indication}
\newacronym{rss}{RSS}{responder sector sweep}
\newacronym{rts}{RTS}{request to send}
\newacronym{rwpm}{RWPM}{random waypoint model}
\newacronym{rx}{RX}{receive}
\newacronym{sc}{SC}{single carrier}
\newacronym{simo}{SIMO}{single-input and multiple-output}
\newacronym{sinr}{SINR}{signal-to-interference-plus-noise ratio}
\newacronym{siso}{SISO}{single-input and single-output}
\newacronym{sls}{SLS}{sector level sweep}
\newacronym{sp}{SP}{service period}
\newacronym{ssw}{SSW}{sector sweep}
\newacronym{ss}{SS}{sector sweep}
\newacronym{sta}{STA}{station}
\newacronym{stf}{STF}{short training field}
\newacronym{tdma}{TDMA}{time division multiple access}
\newacronym{tg3c}{TG3c}{task group IEEE 802.15.3c}
\newacronym{tgad}{TGad}{task group IEEE 802.11ad}
\newacronym{trn}{TRN}{training}
\newacronym{tx}{TX}{transmit}
\newacronym{ula}{ULA}{uniform linear array}
\newacronym{usa}{USA}{uniform square array}
\newacronym{wlan}{WLAN}{wireless local area network}
\newacronym{wpan}{WPAN}{wireless personal area network}

\begin{document}
\title{Sensor Assisted Movement Identification and Prediction for Beamformed 60 GHz Links: A Report}
\author{\IEEEauthorblockN{A.W. Doff, K. Chandra and R. Venkatesha Prasad\\}
        \IEEEauthorblockA{Faculty of Electrical Engineering, Mathematics and Computer Science\\ 
        Delft University of Technology, The Netherlands\\
        Email: a.w.doff@student.tudelft.nl, k.chandra@tudelft.nl, rvprasad@ieee.org}
}
\maketitle
\begin{abstract}
Large available bandwidth in 60\,GHz band promises very high data rates -- in the order of Gb/s.
However, high free-space path loss makes it necessary to employ beamforming capable directional antennas. 
When beamforming is used, the links are sensitive to misalignment in antenna directionality because of movement of devices.
To identify and circumvent the misalignments, we propose to use the motion sensors (i.e., accelerometer and gyroscope) which are already present in most of the modern mobile devices. 
By finding the extent of misaligned beams, corrective actions are carried out to reconfigure the antennas. 
Motion sensors on mobile devices provide means to estimate the extent of misalignments.
We collected real data from motion sensors and steer the beams appropriately. 
The results from our study show that the sensors are capable of detecting the cause of errors as translational or rotational movements.
Furthermore it is also shown that the sensor data can be used to predict the next location of the user. 
This can be used to reconfigure the directional antenna to switch the antenna beam directions and hence avoid frequent link disruptions. 
This decreases the number of beam searches thus lowering the MAC overhead.
\end{abstract}

\section{Introduction} \label{sec:introduction}
A lot of interest has been shown in the 60\,GHz band since the \gls{fcc} has allocated 7\,GHz in the 57-64\,GHz band for unlicensed use~\cite{yong2007,fcc2010,yong2011,TSR60GHz,vp1,vp2}. 
The IEEE has already drafted two standards specifying PHY and MAC layer for short range communication in 60\,GHz band, \textit{viz.}, IEEE 802.15.3c and IEEE 802.11ad~\cite{ieee.3c2009,ieee.ad2012}. 
IEEE 802.15.3c specify PHY and MAC parameters for 60\,GHz \gls{wpan} while IEEE 802.11ad also targets \gls{wlan} services in 60\,GHz frequency band which is back compatible with 2.4/5\,GHz IEEE 802.11b/g/n/ac. 
Moreover, it is also expected that 60\,GHz communication technology will play an important role in the 5G communication scenarios~\cite{5G1,5G2}.


To mitigate the high free space pathloss in 60\,GHz band, directional antennas are used.
Smaller wavelength at 60\,GHz allows the use of antenna array, which takes less space. For example with a half wavelength spacing, up to 16 antennas can be used in a 1\,cm$^2$ \gls{usa}.
Antenna arrays can be efficiently used to form narrow beams to focus signal power in desired direction and also for beam steering. 

However, device movement in 60\,GHz networks can limit the maximum achievable link quality~\cite{surveyVP60GHz,beamswitching_xeuli,iee:toward}. The first cause is \textit{linear motion}, also referred to as a translational movement, which is sensed by variations in the accelerometer data. 
Secondly, \textit{rotational movement} is caused when the user is turning around or changing the orientation of the mobile device.
This can be sensed by variations in the gyroscope and accelerometer sensor data. 
Lastly, \textit{blockage} occurs when the links are interrupted by other users or obstacles while the user is moving around. 
In this case \gls{nlos} paths have to be used.

It is usually difficult to determine the cause of link degradation. 
In~\cite{tsang2011_2} the rate at which the received signal power changes is monitored to identify the error type affecting the link. 
It is important to note that the solution to reclaim the lost link because of one of these errors is not compatible with the other errors. 
Each error requires different compensation. 
For example, in a traditional setup if the device is moving in a linear direction, the \gls{sta}/\gls{ap} does not know what causes the link degradation.
This means it might switch to a \gls{nlos} beam-pair, which has a worse \gls{sinr}.
Hence it is important to correctly identify the communication errors before proceeding to solve them. 
This paper proposes to use the motion sensors, which are already embedded into most modern devices, to help with the identification of the error.

When the device employs its sensors, it can estimate using simple classification techniques if the error is caused by translation, rotation or beam blockage.
Once the error is identified, it is possible to go a step further and also try to resolve this error.
This means that we can predict the next beam-pair and shift to them before the link quality drops significantly. 
If the link quality drops too much, the beamformed link has to be re-established. 
To re-establish the communication link, \gls{tx} and \gls{rx} stations have to restart searching for the best beam pair; we refer to this as \textit{re-beamforming}.
However, with the help of sensor data, movement of devices can be predicted and appropriate antenna weight vectors can be calculated to maintain a stable link. 
So the overheads of frequent beam searching procedure can be avoided. Thus this paper mainly solves two problems: (i)~determining the error that caused the link degradation; and (ii)~predict the next beam-pair such that the link quality remains stable while minimizing the number of re-beamforming procedures.

The rest of this paper is organised as follows.
%
In Section~\ref{sec:network_setup} we describe the architecture of the IEEE 802.11ad to position relevance of our work. Following that, in Section~\ref{sec:sensors_and_movement} we discuss the sensors used and generalize major movements that occur. Section~\ref{sec:id_and_pred} will describe the movement identification algorithm and prediction methods used.
In Section~\ref{sec:test_setup} we describe the test setup and the simulation environment followed by results and discussion in Section~\ref{sec:results}.
Finally we conclude with future outlook in Section~\ref{sec:conclusion}.

\section{Typical 60\,GHz Network} \label{sec:network_setup}
\begin{figure}[h]
	\centering
	\includegraphics[width=0.4\textwidth]{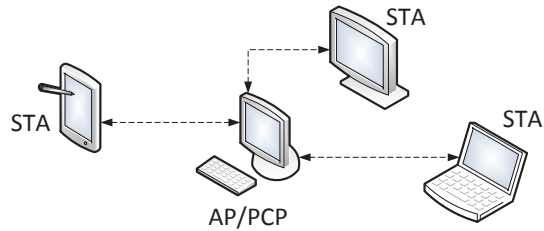}
	\caption{An example of the IEEE 802.11ad \gls{wlan} architecture.}
	\label{fig:WLAN_architecture}
\end{figure}

Similar to the IEEE 802.11b/g/n \gls{bss}, IEEE 802.11ad uses a \gls{pbss}, which is the operating area of IEEE 802.11ad networks.
To provide basic timing to the \glspl{sta}, one \gls{sta} assumes the role of \gls{pcp}, as shown in Fig.~\ref{fig:WLAN_architecture}.

To select the best beam pair for transmit and receive antenna array, IEEE 802.11ad employs an improved variant of the multi-level beamforming scheme proposed in IEEE 802.15.3c~\cite{ieee.3c2009,wang2009}.

The IEEE 802.11ad beamforming process consists of two phases.
The first phase is the \gls{sls}.
Its purpose is to allow communications between two \glspl{sta}.
The \gls{sls} is followed by the \gls{brp}.
In general the \gls{brp} is used to further train \gls{rx} and \gls{tx} antennas of a \gls{sta} after the \gls{sls} phase.
This phase is a request/response based process.

The \gls{sta} requesting beamforming is referred to as the initiator, while the receiving \gls{sta} is referred to as the responder.
In order to further track the beams/channel a \gls{bt} phase can be used after the \gls{sls}/\gls{brp}.

\begin{figure}[h]
	\centering
	\includegraphics[width=0.4\textwidth]{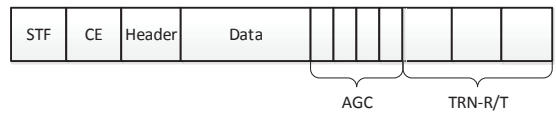}
	\caption{The IEEE 802.11ad \gls{brp} packet structure.}
	\label{fig:PHYpacket_ad}
\end{figure}

The beamforming process in the IEEE 802.11ad standard offers the option to add additional \gls{trn} bits at the end of a packet.
This is illustrated in Fig.~\ref{fig:PHYpacket_ad} where optional training bits can be appended to a data packet.
The \gls{agc} fields are added in order to account for the variation in signal strength when transmitting and receiving beam \gls{trn} fields. The \gls{bt} phase is an essential part in maintaining link quality. However, difficulties arise when movement is involved.
If the user moves, the error has to be identified and the beam-pairs will have to adapt accordingly.

\section{Sensors and Movement} \label{sec:sensors_and_movement}
Sensors are an intricate part of this work, thus we discuss them briefly in this section followed by two types of movements mentioned earlier.

\subsection{Sensors}
A smartphones has many sensors, among which \textit{accelerometers}, \textit{gyroscopes} and \textit{magnetometers} can be seen as the most relevant to detect motion. Gyroscopes measure angular velocity in rad/s based on the Coriolis force while accelerometer measures the linear acceleration in \,m/s$^2$. The magnetometer can be used as a digital compass.
Both the accelerometer and the gyroscope data are used in the error identification phase and will be represented by $\vect{a} = [a_x, a_y, a_z]$ and $\vect{g} = [g_x, g_y, g_z]$ respectively.

To retrieve useful information from these sensors it is possible to combine the data from two or more sensors.
The combination of sensors is referred to as a virtual sensor.
The \textit{rotation vector sensor} is such a virtual sensor, where accelerometer, gyroscope and  magnetometer data are fused. 
The rotation vector sensor gives the orientation of the device relative to the East-North-Up coordinates, and is represented as $\vect{\epsilon} = [\phi, \theta, \Psi]$.
The azimuth angle $\phi$ will be used in the movement prediction phase to determine the direction of movement.

\subsection{Types of movements}
\begin{figure}[h]
	\centering
	\includegraphics[width=0.4\textwidth]{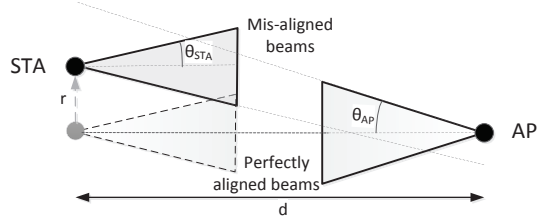}
	\caption{A movement in a linear direction.}
	\label{fig:translational_beam_movement}
		\vspace{-5mm}
\end{figure}

When \gls{sta} and \gls{ap} are connected, we assume their beams are aligned. However, translational or rotational movements can cause link degradation.
If the \gls{sta} starts to move in a linear direction, both the \gls{sta} and the \gls{ap} need to change beam direction, as can be seen in Fig.~\ref{fig:translational_beam_movement}. 
This means that if the \gls{sta} starts to move, it needs to notify the \gls{ap} such that both devices know that their beam-pairs need to be realigned.

\begin{figure}[h]
	\centering
	\includegraphics[width=0.4\textwidth]{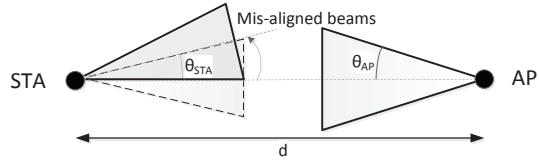}
	\caption{An angular movement.}
	\label{fig:rotational_beam_movement}
		\vspace{-5mm}
\end{figure}

When the \gls{sta} is turning we only need to change the beam direction of the \gls{rx} \gls{sta}, as can be seen in \ref{fig:rotational_beam_movement}.
It can be easily seen that the \gls{sta} should switch its beam if it has rotated more than $\theta_{STA}$.

To get an understanding of how often beam switching needs to be done, angular speeds were measured~\cite{tsang2011_2}.
\begin{table}[h]
	\centering
	\caption{Angular displacement of a smartphone under different activities.}
	\begin{tabular}{p{0.25\textwidth} | p{0.15\textwidth}}
		\hline
		Activities & Angular displacement in 100\,ms \\ \hline\hline
		Reading, web browsing (no change of orientations) & $6^\circ - 11^\circ$ \\ \hline
		Reading, web browsing (horizontal from/to vertical orientation changes) & $30^\circ - 36^\circ$ \\ \hline
		Playing games & $72^\circ - 80^\circ$ \\ \hline
	\end{tabular}
	\label{table:rotational_movement}
\end{table}
Table~\ref{table:rotational_movement} shows the necessity of performing frequent re-beamforming due to rotations if no additional information is known.
This was also explored and shown in~\cite{park2010} where beam switching needed to be done every 14-54\,ms during rotations, depending on the antenna setup.
\section{Identifying and predicting movements} \label{sec:id_and_pred}
The first objective is to identify what caused the error: translational movement, rotational movement or blockage.
The secondary objective is to predict the next beam sector/beam-pair when we know the origin of the error.
\subsection{Identify error}
The first phase is to identify the error by performing activity recognition to detect if the user is standing still, moving straight, turning or both turning and moving. 
A simple, yet very effective algorithm used in activity recognition is \gls{knn}~\cite{peterson2009} which uses a feature vector to identify features specific to a certain activity.

Let $k$ be the number of nearest neighbours and $\vect{T} = \{\vect{x}_1, \vect{x}_2,\dots,\vect{x}_N\}$ be the training samples.
$\vect{x}_i = (\vect{f}_i, c_i)$, where $\vect{f}_i$ is the feature vector of the training samples $\vect{x}_i$ and $c_i$ is the class that $\vect{x}_i$ belongs to. 
The feature vector was chosen to contain the mean and standard deviation of $\vect{a}$ and $\vect{g}$, as well as the maximal autocorrelation of $\vect{a}$ to better detect the steps taken by a user. 
Thus $\vect{f} = [\mu_{\vect{a}}, \mu_{\vect{g}}, \sigma_{\vect{a}}, \sigma_{\vect{g}}, \text{max}(R_{\vect{a}})]$.

A new sample $\vect{\hat{x}} = (\vect{\hat{f}},\hat{c})$ is classified as shown in ALGO.~\ref{pseudocode:phase1}.
\begin{algorithm}
	\begin{algorithmic}[1]
		\FOR {\textbf{each} $\vect{\hat{x}} = (\vect{\hat{f}},\hat{c})$}
		\STATE {Calculate the distance $d(\vect{\hat{f}},\vect{f}_i)$ between $\vect{\hat{x}}$ and all $\vect{x}_i$ in $\vect{T}$.}
		\STATE {Sort $\vect{T}$ ascending based on the distance $d(\vect{\hat{f}},\vect{f}_i)$.}
		\STATE {Select the first $k$ samples from $\vect{T}$, these are the $k$ points closest to $\vect{\hat{x}}$.}
		\STATE {Assign a class to $\hat{c}$ based on the majority vote of the $k$ classes.}
		\ENDFOR
	\end{algorithmic}
	\caption{\gls{knn}}
	\label{pseudocode:phase1}
\end{algorithm}
Multiple measures of distances can be used to calculate the distance $d(\vect{\hat{f}},\vect{f}_i)$ between the feature vector of the input sample and the training samples. 
We chose to use the $l^2$ norm, where the distance $d$ is calculated as $d(\vect{\hat{f}},\vect{f}) = \sqrt{ \sum_{j=1}^n|\hat{f}_j-f_j|^2}$. Here $n = 15$ is the length of the feature vector.

Blockage can also indirectly be detected using sensors. 
If the \gls{sinr} drops and the device does not measure any movement, it is apparent that the beam was blocked. 
Note that this conclusion is only valid if we assume movement and blockage do not occur at the same time. 
If they do occur at the same time, it will be difficult to identify the source of the error.

\subsection{Movement prediction}
The second phase is to predict the next best beam-pair based on the measurements from the sensors. Three prediction methods are investigated.

\subsubsection{No prediction}
This prediction method simply means every time the signal power drops, beam searching needs to be performed.

\subsubsection{Simple prediction}
The second prediction method is done by extrapolating the next beam sector from the previous beam sector.
If we take $\vect{S_c} = [x_c, y_c]^T$ as the current sector and $\vect{S_p} = [x_p, y_p]^T$ as the previous sector.
Then the next sector $\vect{S_n}$ is calculated as
\begin{equation*}
    \vect{S_n} = \begin{bmatrix} x_n \\ y_n \end{bmatrix} = \begin{bmatrix} x_c + \sgn(x_c - x_p) \\ y_c + \sgn(y_c - y_p) \end{bmatrix},
\end{equation*}
where $\sgn(*)$ is the signum operator.
As an example, if the user moves from beam sector $[x_p, y_p]^T = [3, 1]^T$ to $[x_c,y_c]^T = [3,2]^T$, the next beam sector is predicted to be $[x_n,y_n]^T = [3,3]^T$.
This method works well if the user is always walking in one direction, however if turns are made this method will wrongly predict the next beam-pair.

\subsubsection{Sensor prediction}
The third and last prediction method is by using the rotation vector sensor in the device.
The rotation vector sensor gives the orientation of the device.
It is assumed the azimuth angle from this sensor can be used as an indication of the direction of the user.
The prediction of the next beam sector $\vect{S_n}$ is calculated from this direction as follows:
\begin{equation*}
	\vect{S_n} = \begin{bmatrix} x_n \\ y_n \end{bmatrix} = \begin{bmatrix} x_c + \nint(\sin{\phi}) \\ y_c + \nint(\cos{\phi}) \end{bmatrix},
\end{equation*}
where $\nint(*)$ is the nearest integer, or round function and $\phi$ represents the azimuth angle of the device orientation.

\section{Test setup} \label{sec:test_setup}
For the prediction phase, without loss of generality, the following example scenario is assumed.
A user device (\gls{sta}) is connected to a 60\,GHz \gls{ap} located at the centre of the ceiling. 
The user starts to move, which means the directional beam of the \gls{ap} is no longer aligned with the user.
A dip in signal power is observed and as such the \gls{ap} will need to switch its beam to another direction.
To find the correct beam-pair, re-beamforming is often employed~\cite{park2010}.
Re-beamforming often consumes a significant amount of time to align the beam pairs, which reduces the channel usage for transmitting data.
Thus our goal is to minimize the number of re-beamformings.

\begin{figure}[h]
	\centering
	\begin{subfigure}[h]{0.2\textwidth}
	\includegraphics[width=\textwidth]{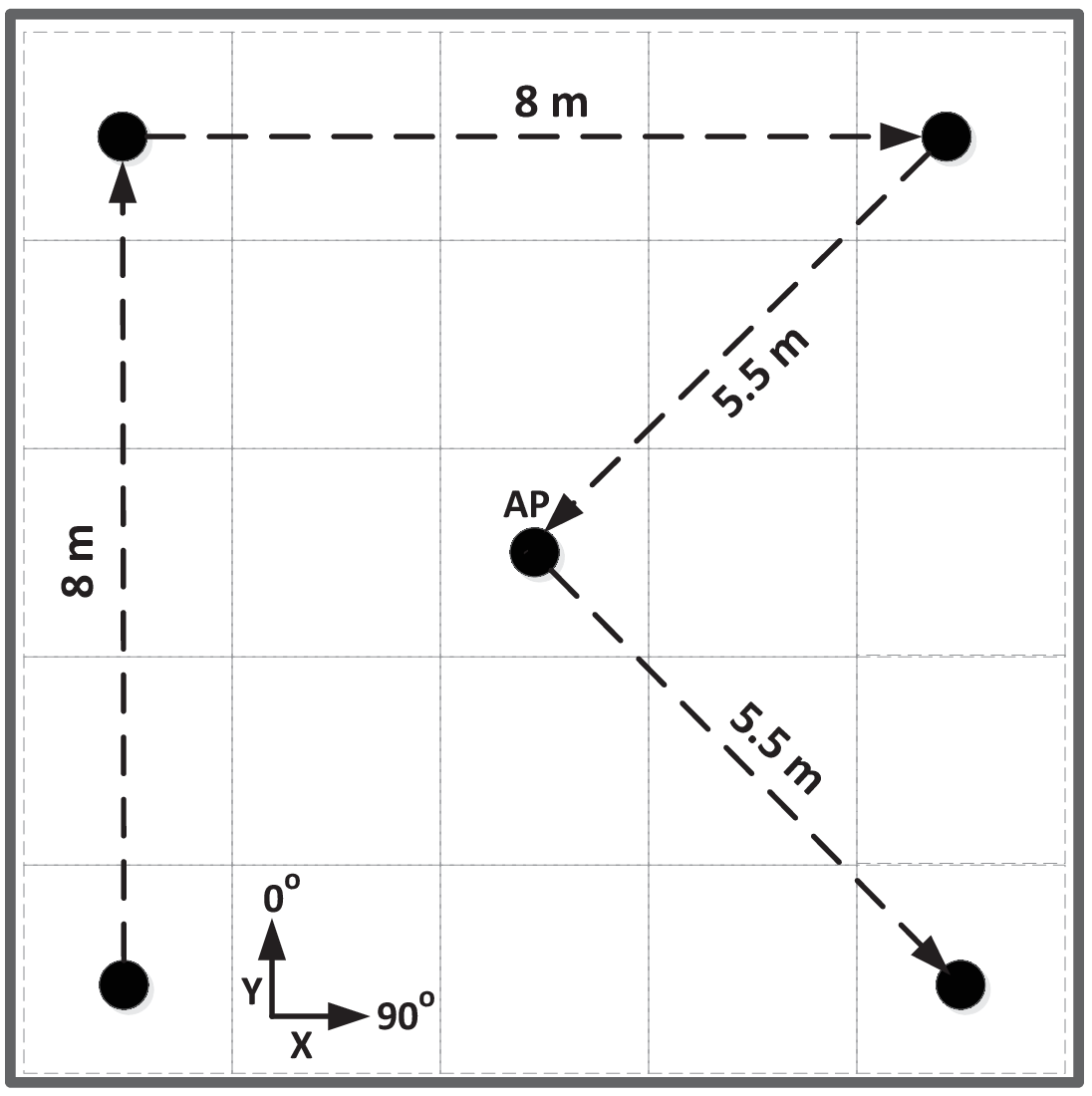}
		\caption{}
		\label{fig:simulation_example_route}
	\end{subfigure}
	\quad
	\begin{subfigure}[h]{0.23\textwidth}
		\includegraphics[width=\textwidth]{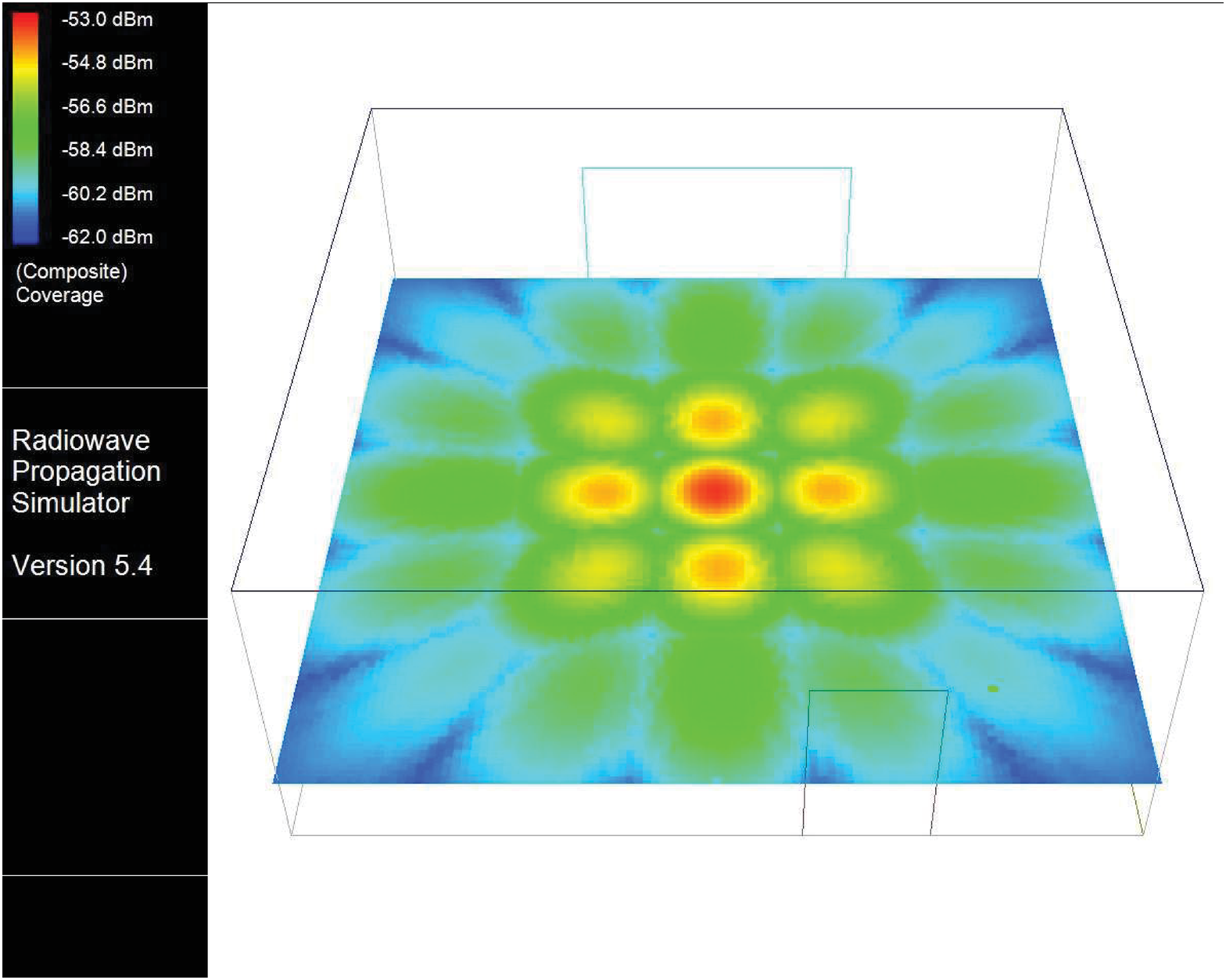}
		\caption{}
		\label{fig:RPS_room_coverage}
	\end{subfigure}
		\vspace{-4mm}
	\caption{The test setup in the room, where (a) shows the $5\times5$ grid and the route and (b) the received power.}
	\label{fig:RPS_room_coverageandroute}
		\vspace{-3mm}
\end{figure}

In order to simulate the 60\,GHz network with mobility we used two stages. 
The first stage is to gather data from a simulated environment using a verified radio frequency propagation simulator, called \gls{rps}~\cite{rps2008}, which provides close to real 60\,GHz signal strength at various locations on a floor plan. 
The second stage consists of collecting \textit{real} experimental sensor data from the user assuming the user is moving along the path shown in \ref{fig:simulation_example_route} using the motion sensors in a Samsung Galaxy SIII smartphone.

In \gls{rps} a room is created with material properties of glass, concrete and wood as shown in \ref{table:dielectric_properties}.
\begin{table}[h]
	\centering
	\caption{Dielectric properties of materials in 60\,GHz~\cite{genc2010}.}
	\begin{tabular}{p{0.13\textwidth} | p{0.10\textwidth} | p{0.05\textwidth} | p{0.05\textwidth}}
		\hline
	    Material & Thickness (m) & $\epsilon_{\text{Re}}$ & $\epsilon_{\text{Im}}$\\ \hline \hline
	    Concrete wall/ceiling & 0.3 & 6.14 & -0.3015\\ \hline
	    Wooden floor & 0.2 & 2.81 & -0.0964\\ \hline
	    Wooden door & 0.04 & 2.81 & -0.0964\\ \hline
	    Glass window & 0.02 & 4.58 & -0.0458\\ \hline
	\end{tabular}
	\label{table:dielectric_properties}
\end{table}
A $100\times100$ grid is placed at a height of 1.5\,m to simulate possible positions and height of users when holding a mobile device.
\Gls{rps} is able to measure the signal power at every position in this grid.
A directional \gls{tx} antenna is placed in the middle of the room at a height of 4\,m, which will act as the \gls{ap}.
This antenna can be directed at 25 different sectors, such that a $5\times5$ grid is created as shown in Fig.~\ref{fig:RPS_room_coverageandroute}.
The \gls{tx} antenna pattern is Gaussian as is also assumed in the \gls{tgad} channel model~\cite{maltsev2010}.

The \gls{rx} antenna pattern of the device is set to be omni-directional. 
This simplifies our analysis because only the \gls{ap} needs to switch its beam in order to maintain a stable link.
Parameters for the \gls{rps} setup are shown in \ref{table:rps_parameters}.
\begin{table}[h]
	\centering
	\caption{\Gls{rps} parameters.}
	\begin{tabular}{p{0.20\textwidth} | p{0.15\textwidth}}
		\hline
		Room dimensions & $10 \times 10 \times 4$\,m\\ \hline
		Carrier frequency & 60\,GHz \\ \hline
		\Gls{tx} \gls{hpbw} & $30^\circ$ \\ \hline
		\Gls{tx} power & 10\,dBm \\ \hline
		\Gls{tx} antenna gain & 14\,dB \\ \hline
		\gls{rx} antenna gain & 0\,dB \\ \hline
		Noise figure & 10\,dB \\ \hline
		Antenna polarization & Left hand circular \\ \hline
	\end{tabular}
	\label{table:rps_parameters}
\end{table}

To predict the next beam sector, it is assumed that the direction of the user with respect to the \gls{ap} can be measured directly using the azimuth angle of the device given by the rotation vector sensor. 
This is of course an abstraction of reality where the user may hold the device in a tilted manner.
There are methods that deal with this problem such that an estimate of the direction of the user can still be found~\cite{roy2014}, however these methods were not implemented here. 
The sensors were all set to the fastest sampling rate specified in the data sheet (100\,Hz for accelerometer and 200\,Hz for gyroscope and orientation sample collection).
%
The sensor data from the user and the \gls{rssi} values from \gls{rps} are combined in Matlab.

To identify the error using the \gls{knn} classifier we choose mean and standard deviation of the accelerometer and gyroscope sensor values as features which have shown good results as reported in~\cite{banos2014}. The autocorrelation of the accelerometer data was also used as a feature to detect the steps taken by a user.

To increase the scale of the experiments with respect to a single simple route, a \gls{rwpm} was used such that statistical analysis on the different prediction methods can be done. 
In the \gls{rwpm} the orientation of the device is assumed to be known.

\section{Experimental Results} \label{sec:results}
\subsection{Identifying errors}
\begin{figure}[h]
	\centering
	\includegraphics[width=0.30\textwidth]{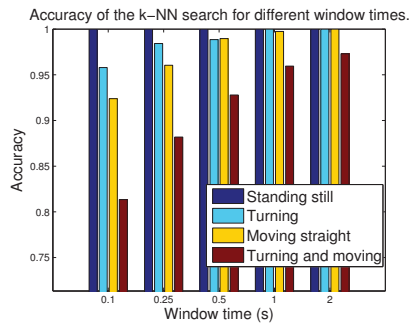}
	\caption{The accuracy of the \gls{knn} search for different window sizes.}
	\label{fig:matlab_statistics_activity_recognition}
	\vspace{-4mm}
\end{figure}

The activity recognition was done by sampling the sensors for 10\,min while doing the activities: (i)~standing still, (ii)~turning, (iii)~moving straight or (iv)~both turning and moving. 
The training vectors were generated from the first 5\,min of the recorded samples.
For the last 5\,min a \gls{knn} search, with $k = 3$ was applied. The accuracy of the \gls{knn} search for different window sizes can be seen in Fig.~\ref{fig:matlab_statistics_activity_recognition}.
A window size of 100\,ms means we are trying to detect the activity from the last 100\,ms.
\begin{figure*}
 \centering
 \begin{subfigure}[h]{0.30\textwidth}
  \includegraphics[width=\textwidth]{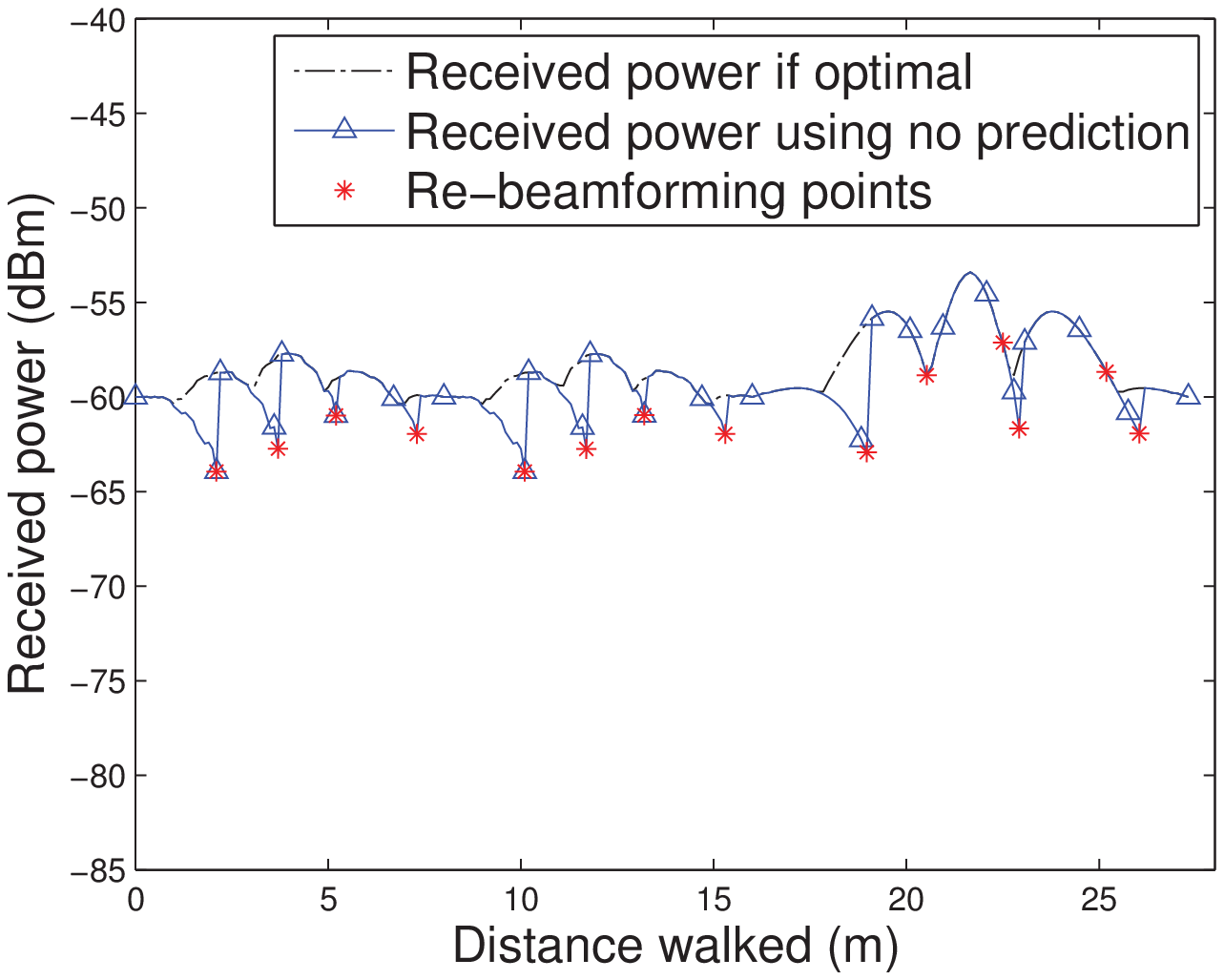}
  \caption{No prediction.}
  \label{fig:matlab_route_no_prediction}
 \end{subfigure}
 \quad
 \begin{subfigure}[h]{0.30\textwidth}
  \includegraphics[width=\textwidth]{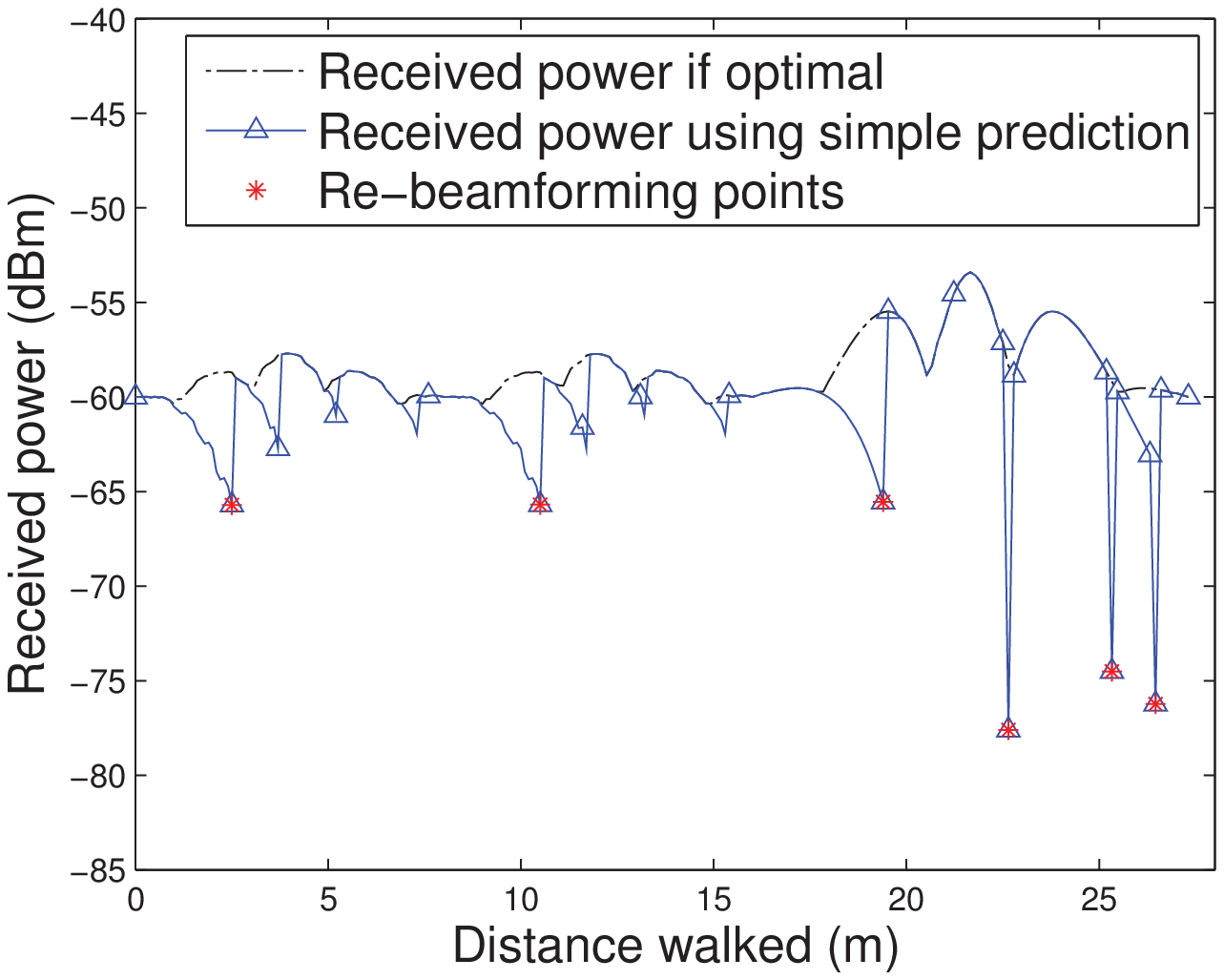}
  \caption{Simple prediction.}
  \label{fig:matlab_route_simple_prediction}
 \end{subfigure}
 \quad
 \begin{subfigure}[h]{0.30\textwidth}
  \includegraphics[width=\textwidth]{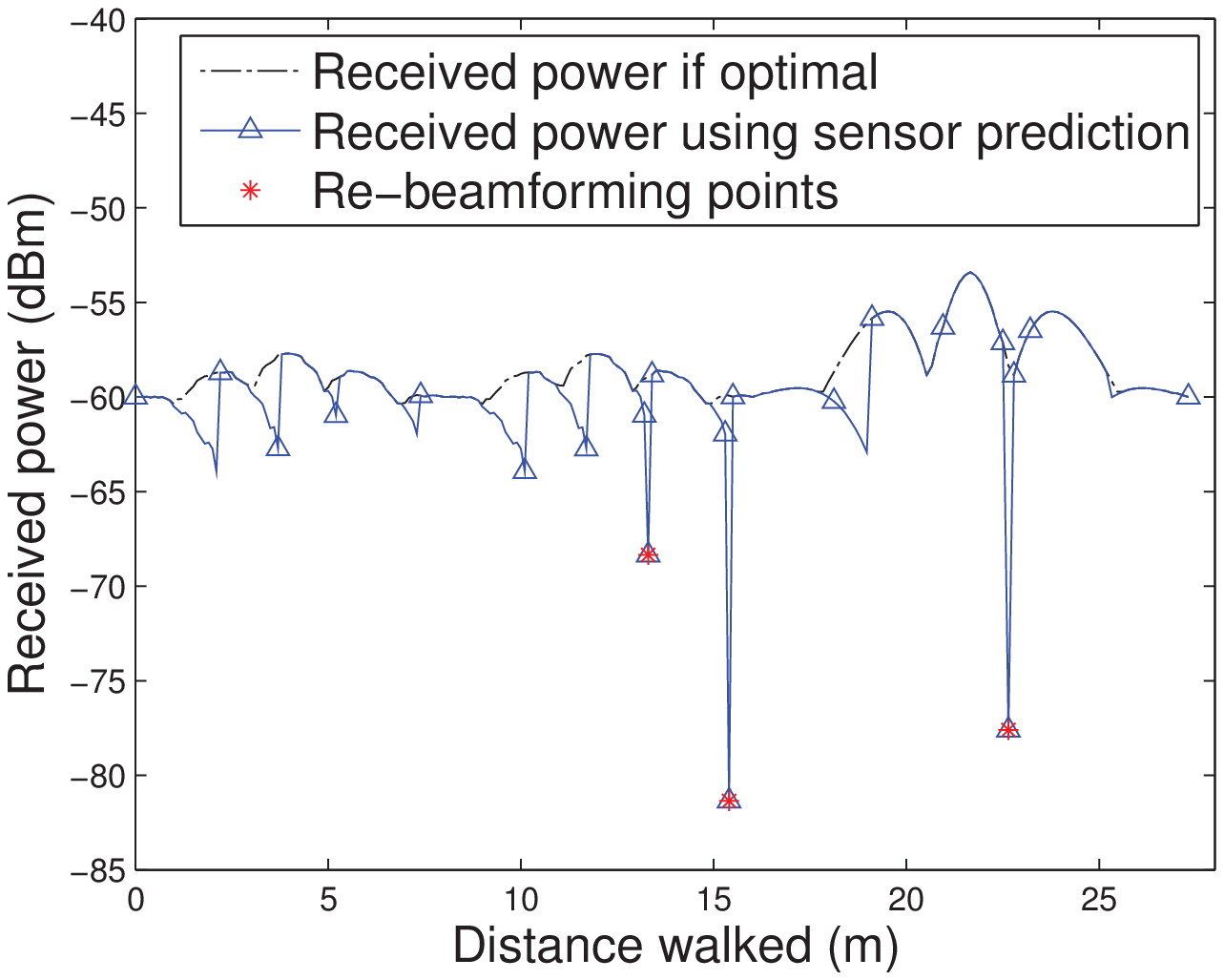}
  \caption{Prediction using sensors.}
  \label{fig:matlab_route_sensor_prediction}
 \end{subfigure}
  	\vspace{-2mm}
 \caption{The received power along the route of \ref{fig:simulation_example_route} using $P_{Dth} = -3$\,dBm and $P_{Rth} = -65$\,dBm.}
 \label{fig:route_received_power}
 	\vspace{-5mm}
\end{figure*}
From Fig.~\ref{fig:matlab_statistics_activity_recognition} it can be seen that an accuracy of 100\% can be achieved when the user is standing still.
This means we were always able to identify if the user was standing still.
Turning could be identified with an accuracy of 96-100\%.
Moving straight was detected with an accuracy of 92-100\%. 
It can be observed that the combination of turning and moving was the hardest to detect with only 81\% to 97\% accuracy.
As was also explored in~\cite{banos2014} the window size has a big impact on the results. 
Basically there is a trade-off between accuracy and speed.
When the window size decreases, the activity will be recognized faster, however the results will be less accurate and vice versa. 

From Fig.~\ref{table:rotational_movement} we observe that angular movement can happen rapidly. 
This means that the window size of 100\,ms is not enough when the movement is faster.
The accelerometer is sampling at roughly 100\,Hz, which means 10 samples are taken every 100\,ms.
To further increase the reaction time for activity recognition, we need to decrease the window size to 10\,ms.
Meaning we would observe barely 1 sample in a window, which is not sufficient to recognize an activity.
To overcome this limitation the sampling rate of the sensors needs to be increased.
If the sampling rate is increased, faster and more accurate results can be obtained.

\subsection{Movement prediction}
To understand the impact of movement prediction, a route was specified as shown in Fig.~\ref{fig:simulation_example_route}.
This route does not contain any obstacles.
Every prediction method is activated only when the received power drops by a certain threshold $P_{Dth}$ -- the drop-off threshold.
A forced re-beamforming is done if the received power reaches the re-beamforming threshold $P_{Rth}$.
For this scenario we chose $P_{Dth} = -3$\,dBm and $P_{Rth} = -65$\,dBm.

The received power along the route of Fig.~\ref{fig:simulation_example_route} for the three prediction methods is shown in Fig.~\ref{fig:route_received_power}.
Re-beamforming needs to be done if a switch is made to a wrong beam sector or if the re-beamforming threshold is reached.
These locations are indicated with a red star in Fig.~\ref{fig:matlab_route_no_prediction}, Fig.~\ref{fig:matlab_route_simple_prediction} and Fig.~\ref{fig:matlab_route_sensor_prediction}.

Using no prediction as seen in Fig.~\ref{fig:matlab_route_no_prediction}, re-beamforming is used every time the signal power drops by more than 3\,dBm.
This results in re-beamforming having to be performed 14 times along the route.
In contrast, Fig.~\ref{fig:matlab_route_simple_prediction} shows that using a simple prediction method, only 6 re-beamformings are needed along the route.
The number of re-beamforming along the route is reduced to 3 if the sensor prediction method is used.

It is not always possible to predict and switch the beam to better beam-pair in terms of received power. This causes sudden drop in power as shown in Fig.\ref{fig:matlab_route_simple_prediction} and Fig.~\ref{fig:matlab_route_sensor_prediction}. 
The sudden drop in signal power is undesired because it forces re-beamforming. Thus in order to minimize this signal power dip it is advised that the prediction methods incorporate a simple beam-pair test to see if the predicted beam-pair actually increases the link quality.

\subsection{Simulation with \gls{rwpm}}
\begin{figure}
	\centering
		\includegraphics[width=0.35\textwidth]{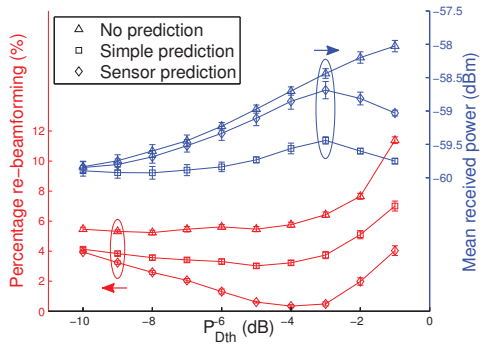}
	\caption{The percentage of re-beamforming and mean received power v/s the drop-off threshold $P_{Dth}$ in the \gls{rwpm}, for the three prediction methods with $P_{Rth} = -65$\,dBm.}
	\label{fig:matlab_statistics}
	\vspace{-6mm}
\end{figure}

To analyse the improvement due to sensor prediction on the number of re-beamforming we used the \gls{rwpm} to simulate the user movement.
100 waypoints were randomly generated in the room and for different drop-off thresholds the number of re-beamformings was calculated.
This was repeated 100 times to obtain a statistically significant mean and standard deviation as shown in \ref{fig:matlab_statistics}.
At every step along the route it is evaluated if re-beamforming is needed.
We use the distance of the route to normalize the number of re-beamforming, such that a percentage of re-beamforming is obtained.
From Fig.~\ref{fig:matlab_statistics} it can be seen that when using sensor prediction and $P_{Dth} =-4$\,dB, 0.5\% re-beamforming is needed.
With no prediction the re-beamforming increases to 6\% for the same $P_{Dth}$. This means the overhead due to re-beamforming is up to 12 times lower if sensors are used under the condition that the \gls{rx} device is omni-directional.
The gain is expected to increase if directional antennas are used. 

The minimum re-beamforming percentage of 0.5\% is obtained at $P_{Dth} = -4$\,dBm using sensor based prediction. 
This is because, for higher drop-off thresholds ($P_{Dth} = -3$\,dBm to $-1$\,dBm) beam switching is triggered too often, increasing the re-beamforming percentage. 
The increase in re-beamforming for lower drop-off thresholds ($P_{Dth} = -10$\,dBm to $-5$\,dBm) can be attributed to two reasons.
First, as $P_{Dth}$ is decreased, the forced re-beamforming threshold $P_{Rth}$ will be reached more often resulting in triggering frequent re-beamforming.
Second, the low $P_{Dth}$ causes the \gls{ap} to wait too long to switch beams.
If it takes too long to switch beams based on the prediction, the predicted beam is more likely to be misaligned, which leads to re-beamforming.

The trade-off can be seen when looking at the mean received power along the route as shown in Fig.~\ref{fig:matlab_statistics}.
If sensor prediction is employed, the mis-predictions cause the mean received power to be lower compared to no prediction.
Still, for $P_{Dth} = -10$\,dBm to $-3$\,dBm the difference in mean received power between sensor prediction and no prediction is no more than 0.5\,dBm.

In Fig.~\ref{fig:matlab_statistics} we see it is possible to obtain a favorable $P_{Dth}$ in terms of both re-beamformings and received power.
Using sensor prediciton, the mean received power is maximized at $P_{Dth} = -3$\,dBm and the re-beamforming percentage is minimized at $P_{Dth} = -4$\,dBm.

\section{Conclusion} \label{sec:conclusion}
User movement introduces translational, rotational or blockage errors in 60\,GHz networks which can cause beam mis-alignment when directional antennas are used. 
Using commonly available sensors in the mobile devices, it is possible to predict the movement and quickly realign the beams without disrupting the connection. 
This work takes the first steps to incorporate sensor data as a means of improving network performance. To identify the error, \gls{knn} classification was successfully used, however the drawback of \gls{knn} is the need for training. Currently the IEEE 802.11ad standard has no field wherein the sensor data can be transmitted. 
This means either the IEEE 802.11ad dataframe needs to include the sensor data or the sensor data needs to be transmitted in a separate packet.
More investigation is thus required to incorporate our findings. 
The prediction methods in the case of blockage and directional \gls{rx} antennas will also need to be investigated further.

\section*{Acknowledgement}
This work is funded by the Dutch national project SOWICI.

\bibliographystyle{IEEEtran}
\bibliography{IEEEabrv,references}

\begin{thebibliography}{10}
\providecommand{\url}[1]{#1}
\csname url@samestyle\endcsname
\providecommand{\newblock}{\relax}
\providecommand{\bibinfo}[2]{#2}
\providecommand{\BIBentrySTDinterwordspacing}{\spaceskip=0pt\relax}
\providecommand{\BIBentryALTinterwordstretchfactor}{4}
\providecommand{\BIBentryALTinterwordspacing}{\spaceskip=\fontdimen2\font plus
\BIBentryALTinterwordstretchfactor\fontdimen3\font minus
  \fontdimen4\font\relax}
\providecommand{\BIBforeignlanguage}[2]{{%
\expandafter\ifx\csname l@#1\endcsname\relax
\typeout{** WARNING: IEEEtran.bst: No hyphenation pattern has been}%
\typeout{** loaded for the language `#1'. Using the pattern for}%
\typeout{** the default language instead.}%
\else
\language=\csname l@#1\endcsname
\fi
#2}}
\providecommand{\BIBdecl}{\relax}
\BIBdecl

\bibitem{yong2007}
S.~K. Yong and C.-C. Chong, ``An Overview of Multigigabit Wireless Through
  Millimeter Wave Technology: Potentials and Technical Challenges,''
  \emph{EURASIP J. Wirel. Commun. Netw.}, vol. 2007, no.~1, pp. 50--50, Jan.
  2007.

\bibitem{fcc2010}
FCC, ``Code of Federal Regulation, title 47 Telecommunication, chapter 1, part
  15.255,'' October 2010.

\bibitem{yong2011}
S.-K.~S. Yong, P.~Xia, and A.~V. Garcia, \emph{60 GHz Technology for Gbps,
  WLAN, WPAN: From Theory to Practice}.\hskip 1em plus 0.5em minus 0.4em\relax
  John Wiley \& Sons, Ltd, 2011.

\bibitem{TSR60GHz}
T.~Rappaport, J.~Murdock, and F.~Gutierrez, ``State of the Art in 60-GHz
  Integrated Circuits and Systems for Wireless Communications,''
  \emph{Proceedings of the IEEE}, vol.~99, no.~8, pp. 1390--1436, Aug 2011.

\bibitem{vp1}
B.~L. Dang, R.~Prasad, and I.~Niemegeers, ``On the MAC protocols for Radio over
  Fiber indoor networks,'' in \emph{Communications and Electronics, 2006. ICCE
  '06. First International Conference on}, Oct 2006, pp. 112--117.

\bibitem{vp2}
N.~Wangi, R.~Prasad, M.~Jacobsson, and I.~Niemegeers, ``Address
  autoconfiguration in wireless ad hoc networks: protocols and techniques,''
  \emph{Wireless Communications, IEEE}, vol.~15, no.~1, pp. 70--80, February
  2008.

\bibitem{ieee.3c2009}
``IEEE Standard for Information technology - Telecommunications and information
  exchange between systems,'' \emph{IEEE Std 802.15.3c-2009 (Amendment to IEEE
  Std 802.15.3-2003)}, pp. c1--187, Oct 2009.

\bibitem{ieee.ad2012}
``IEEE Standard for Information technology--Telecommunications and information
  exchange between systems,'' \emph{IEEE Std 802.11ad-2012 (Amendment to IEEE
  Std 802.11-2012, as amended by IEEE Std 802.11ae-2012 and IEEE Std
  802.11aa-2012)}, pp. 1--628, 2012.

\bibitem{5G1}
F.~Boccardi, R.~Heath, A.~Lozano, T.~Marzetta, and P.~Popovski, ``Five
  disruptive technology directions for 5G,'' \emph{Communications Magazine,
  IEEE}, vol.~52, no.~2, pp. 74--80, February 2014.

\bibitem{5G2}
T.~Rappaport, S.~Sun, R.~Mayzus, H.~Zhao, Y.~Azar, K.~Wang, G.~Wong, J.~Schulz,
  M.~Samimi, and F.~Gutierrez, ``Millimeter Wave Mobile Communications for 5G
  Cellular: It Will Work!'' \emph{Access, IEEE}, vol.~1, pp. 335--349, 2013.

\bibitem{surveyVP60GHz}
B.~V. Quang, R.~Venkatesha~Prasad, and I.~Niemegeers, ``A Survey on Handoffs -
  Lessons for 60 GHz Based Wireless Systems,'' \emph{Communications Surveys
  Tutorials, IEEE}, vol.~14, no.~1, pp. 64--86, First 2012.

\bibitem{beamswitching_xeuli}
X.~An, C.-S. Sum, R.~Prasad, J.~Wang, Z.~Lan, J.~Wang, R.~Hekmat, H.~Harada,
  and I.~Niemegeers, ``Beam switching support to resolve link-blockage problem
  in 60 GHz WPANs,'' in \emph{Personal, Indoor and Mobile Radio Communications,
  2009 IEEE 20th International Symposium on}, 2009, pp. 390--394.

\bibitem{iee:toward}
B.~L. Dang, R.~V. Prasad, I.~G. Niemegeers, M.~G. Larrode, and A.~M.~J. Koonen,
  ``Toward a Seamless Communication Architecture for In-building Networks at
  the 60 GHz band,'' in \emph{LCN'06}, 2006, pp. 300--307.

\bibitem{tsang2011_2}
Y.~Tsang and A.~Poon, ``Detecting Human Blockage and Device Movement in mmWave
  Communication System,'' in \emph{Global Telecommunications Conference
  (GLOBECOM 2011), 2011 IEEE}, Dec 2011, pp. 1--6.

\bibitem{wang2009}
J.~Wang, Z.~Lan, C.~woo Pyo, T.~Baykas, C.-S. Sum, M.~Rahman, J.~Gao,
  R.~Funada, F.~Kojima, H.~Harada, and S.~Kato, ``Beam codebook based
  beamforming protocol for multi-Gbps millimeter-wave WPAN systems,''
  \emph{Selected Areas in Communications, IEEE Journal on}, vol.~27, no.~8, pp.
  1390--1399, October 2009.

\bibitem{park2010}
M.~Park and H.~K. Pan, ``Effect of Device Mobility and Phased Array Antennas on
  60 GHz Wireless Networks,'' in \emph{Proceedings of the 2010 ACM
  International Workshop on mmWave Communications: From Circuits to Networks},
  ser. mmCom '10, 2010, pp. 51--56.

\bibitem{peterson2009}
L.~E. Peterson, ``{K}-nearest neighbor,'' vol.~4, no.~2, p. 1883, 2009,
  {revision \#136646}.

\bibitem{rps2008}
J.~Deissne and et~al, ``{RPS Radiowave Propagation Simulator User
  Manual-Version 5.4},'' \emph{Actix GmbH}, 2008.

\bibitem{genc2010}
Z.~Genc, U.~Rizvi, E.~Onur, and I.~Niemegeers, ``Robust 60 GHz Indoor
  Connectivity: Is It Possible with Reflections?'' in \emph{Vehicular
  Technology Conference (VTC 2010-Spring), 2010 IEEE 71st}, May 2010, pp. 1--5.

\bibitem{maltsev2010}
A.~Maltsev and et. al., ``Channel Models for 60 GHz WLAN Systems,'' May 2010.

\bibitem{roy2014}
N.~Roy, H.~Wang, and R.~Roy~Choudhury, ``I Am a Smartphone and I Can Tell My
  User's Walking Direction,'' in \emph{Proceedings of the 12th Annual
  International Conference on Mobile Systems, Applications, and Services}, ser.
  MobiSys '14, 2014, pp. 329--342.

\bibitem{banos2014}
O.~Banos, J.-M. Galvez, M.~Damas, H.~Pomares, and I.~Rojas, ``Window Size
  Impact in Human Activity Recognition,'' \emph{Sensors}, vol.~14, no.~4, pp.
  6474--6499, 2014.

\end{thebibliography}
\end{document}